# Comparison between Poissonian and Markovian Primary Traffics in Cognitive Radio Networks

Abdelaali Chaoub[1], Elhassane Ibn-Elhaj[2]

[1] Laboratory of Electronic and Communication, Mohammadia School of Engineers, Mohammed V-Agdal University
Rabat, Morocco
chaoub.abdelaali@gmail.com

[2] Department of Telecommunication, National Institute of Posts and Telecommunications
Rabat, Morocco
*IEEE Member, ibnelhaj@ieee.org*

**Abstract**
Cognitive Radio generates a big interest as a key cost-effective solution for the underutilization of frequency spectrum in legacy communication networks. The objective of this work lies in conducting a performance evaluation of the end-to-end message delivery under both Markovian and Poissonian primary traffics in lossy Cognitive Radio networks. We aim at inferring the most appropriate conditions for an efficient secondary service provision according to the Cognitive Radio network characteristics. Meanwhile, we have performed a general analysis for many still open issues in Cognitive Radio, but at the end only two critical aspects have been considered, namely, the unforeseen primary reclaims in addition to the collided cognitive transmissions due to the Opportunistic Spectrum Sharing. Some graphs, in view of the average Spectral Efficiency, have been computed and plotted to report some comparative results for a given video transmission under the Markovian and the Poissonian primary interruptions.

## 1. Introduction

Last decades have witnessed an explosive growth of emerging technologies and services hugely demanding in terms of Quality of Service (QoS) and bandwidth. As a result, spectral resources have been exhausted and frequency bands have become crowded [1]. Contrarily, actual observations and real statistics taken on some frequency bands underline the low and discontinuous utilization of the spectrum over time and geographical locations [2] [3]. Consequently, legacy frequency allocation process has proven to be inefficient and requires a profound re-evaluation to move towards open frequency allocation policies. To this end, the Cognitive Radio (CR) [4] has been introduced as a promising approach to enhance the spectral efficiency and mitigate the spectrum scarcity and unavailability.

1.1 Preliminary

In the Cognitive Radio context, every telecommunication system will define two categories of spectrum users. The first would be the primary network called Primary Users (PUs), which holds the spectrum license, and the second is the secondary network referred to as Secondary Users (SUs), allowed to use the spectrum holes provided they do not interfere with the licensed users.

The coexistence of both primary and secondary systems within the same architecture has been made possible by exploiting two well regarded concepts, namely, the Negotiated Spectrum Sharing (NSS) and the Opportunistic Spectrum Sharing (OSS) [5] for CR networks.

In the NSS approach, primary users are willing to accept sharing their licensed spectrum with a secondary system for a given monetary cost with respect to a predetermined schedule. This regime needs a deep modification in spectrum regulation because this spectrum leasing must be explicitly agreed upon between the primary and secondary systems.

The OSS [6] is a key component in Cognitive Radio networks and an optimal way to exploit the spectrum holes that has been temporarily vacant by license owners. In fact, periodically, secondary users conduct a local sensing measurement campaigns. The collected measurements are exchanged between different SUs and communicated to a third trusted host, called centralized scheduler [18], to make a global decision about the potential spectrum holes. Unfortunately, there is a penalty for either false alarm or miss detection du to failure in detecting the presence or not of the primary signal. In this sense, the "hidden node" problem is one of the recurring phenomena in spectrum monitoring issues and is still in open field of research [7].

The use of a cooperative sensing system should enable drawing a complete picture of the surrounding primary and secondary users and as a result alleviate the occurrence of this problem. Authors in [8] have given an excellent summary of recent advances in cooperative spectrum sensing. The hidden node issue is beyond the scope of this paper. From the sensing-inferred decisions, the third party assigns to each secondary user a Secondary User Link (SUL) (Fig. 1) formed using a composition of multiple subchannels (SCs) judged as being vacant and temporarily accessible by SUs. Afterwards, each SU starts its transmission through its allocated SUL. The set of chosen subchannels should be scattered over multiple PU frequency bands in order to be able to battle on two crucial challenges: (1) to cap the interferences caused by the primary traffic reclaims below a predefined design specification, and (2) to reduce the number of jammed subchannels once the primary user appears during the lifetime of a SUL. This principle of dispersing the selected subchannels over the frequency domain is referred to as the Spectrum Pooling Concept [9].

To overcome the problem of packet loss incurred by the primary interruptions, we model the lost packets as erasures and we employ some erasure correcting codes [10]. This approach comprises two axes: channel coding and/or source coding. The channel coding is used to compensate for the corrupted packets due to PU appearance and source coding permits recovering the secondary content up to a certain quality commensurate to the number of packets received. In this work, we combine both techniques above-mentioned. In [11], we have suggested to exploit the Joint Source Channel Coding (JSCC) approach based on the Multiple Description Coding.

In summary, the Cognitive Radio solution is a well regarded agile technology for managing, controlling and optimizing the frequency spectrum allocation. It has gained considerable maturity during the last years. This emerging approach promises great future technological advances and has a huge potential to be exploited for enhancing a wide range of traditional networks especially those "quality hungry" such as multimedia and wireless networks.

Multimedia services are a new field that has gained high user acceptance during the last decades. Fundamental issues in this area include many topics especially data compression, coding and transmission. Regarding the increasing demand for more Quality of Service (QoS) and bandwidth, Cognitive Radio networks are expected to shape the future of multimedia services by means of enhancing the multimedia applications with improved end user perceived quality. The tricky part in this topic is tailoring the legacy multimedia services to suit the specificities of the CR context, which renders the problem of studying the multimedia traffic transmission over CR networks exciting [12] [13].

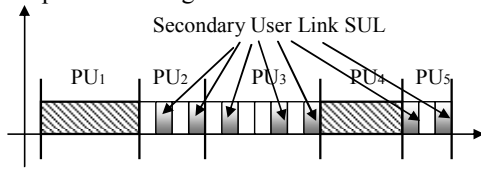

Fig. 1 Spectrum Pooling Concept.

## 1.2 Stochastic Traffic Models

Like any other communication paradigm, stochastic models are required to characterize traffic patterns in Cognitive Radio networks. In deed, the study of interactions between primary and secondary users needs to be further complemented and generalized for many Cognitive Radio contexts. Each traffic type has characteristics that distinguish it from the other types and predicting the traffic behavior is a non-trivial task. In the current contribution, the primary traffic models comprise two classes: Markovian and Poissonian primary traffics. In the Markovian case, we assume that the primary traffic is bursty and has strong correlation properties; the subchannel will remain in a given state for a relatively long duration and do not evolve in a fast way. Such primary traffics access subchannels following a Markovian process [33] [34]. On the other hand, we consider the primary user applications that have a traffic which is dynamic, weakly correlated and varying fast, such primary traffics evolves following a Poissonian process [12] [15].

In literature, we find other primary traffic patterns like the Binomial process [14]. This law makes the following assumptions: One packet is sent per subchannel and then the primary traffic interruptions concern a series of subchannels simultaneously leading to a large variance and a small mean. The primary traffic makes only one attempt to convey a packet across a subchannel and no retry mecanismes are implemented. Corrupted packets are lost forever. This statistical law is a good replacement for the Poissonian process where there are a finite number of sources. In practice, the binomial distribution is more suitable than the Poisson one when $n > 30$ and $np < 5$ or when $n > 50$ and $p < 0.1$ with $p$ the binomial coefficient and $n$ the number of the finite sources.

## 1.3 Literature Review

There exist some research efforts on the problem of secondary traffic transmission over Cognitive Radio networks. In [12], Kushwaha, Xing, Chandramouli and

Heffes have studied the transmission of multimedia traffic (Fig. 2) over CR networks, the primary traffic arrival was modelled as a Poisson process (Fig. 4) and Luby Transform (LT) codes [16] have been used as channel correcting code and also for some coordination reasons. They have proposed a QoS metric to order the available subchannels (SCs) in the decreasing order of their quality to establish the transmission link efficiently. They investigate the spectral efficiency of the selected SUL in terms of successful transmission probability of the required number of packets needed for recovering the original multimedia stream. Unfortunately this paper has neglected the sharing aspect of the network illustrated by the multiple secondary transmissions over the same CR architecture and as a result more packets will be discarded due to collisions which impede considerably the performance of the CR network.

The contribution [15] address the multimedia traffic transmission problem through distributed Cognitive Radio networks using fountain codes under different subchannel selection policies in a fading environment with Poissonian primary traffic reclaims. Additionally, we have proposed a solution consisting of the creation of many secondary user links with high reliability using a specific algorithm that we have introduced to alleviate traffic collisions du to the OSS feature of CR networks.

Some authors like [17] have modelled the subchannel availability using the Binomial process and have suggested the use of a general model for link maintenance to enable the provision of the secondary service even if the SUL get broken due to primary interruptions. The proposed link maintenance model relies on the redundancy principle. Numerical simulations have been performed in terms of Goodput to evaluate the robustness of the proposed model. In [18], we address the problem of video transmission over shared CR networks using progressive compression source coding associated to fountain codes with the assumption that the primary traffic follows the Binomial law. In addition to the dynamics between the primary and secondary networks, the paper [18] considers the conflicts between the secondary users themselves in accessing the common shared frequency bands.

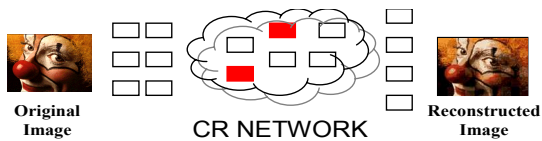

Fig. 2 Distributed multimedia transmission.

Papers like [19] have tried to evaluate the performance of the secondary service under both Binomial and Poissonian primary applications on Cognitive Radio networks and analyze the degree of use of the spectral resource in a lossy context subject to subchannels fading and noise in addition to primary reclaims.

Other studies like [20] and [21] support the Markovian process and have modelled the subchannel (SC) availability using Markov chain for which transition probabilities are known but the state can be only partially observed, so the subchannels to be sensed and accessed constitute a Partially Observable Markov Decision Process (POMDP) [20]. Unfortunately, the studies [20] and [21] propose solutions that are specific to the Markovian model and need to be debated for other traffic patterns.

In [22], Cuiran and Chengshu have investigated the successful transmission over Cognitive Radio networks shared by several SUs using the TDMA technique, they have supposed that the frame includes several slots, each SU transmits in his assigned slot and can transmits in the other slots with certain probability, the results have been presented in terms of throughput and energy efficiency. They have considered that the only reception failure reason would be packet collisions due to time sharing. This study has not taken into account the interference effects caused by the primary user appearance. The reception failure depends also on the Primary traffic type and arrival model which affects the reception of the whole transmitted packets. Furthermore, it may happen that the SU does not transmit data in his own slot because there may be no data to transfer, hence the probability that the secondary device transmits in its assigned slot should be, in practice, less than one.

Studies [33] and [34] look at the problem of distributed multimedia traffic transmission over Cognitive Radio TDMA networks under markovian primary traffic interruptions. A general OSS model has been adopted as a mean of evaluating the secondary conflicts generated by the concurrent access to shared idle subchannels. The paper [34] has introduced a new quality metric to permit a decreasing order of the available subchannels in terms of reliability.

Despite of the fact that there have been many contributions in the Cognitive Radio research area, at our best knowledge no attempt has been made before to report a comparative study between Markovian and Poissonian primary reclaims in terms of Spectral Efficiency evaluation of a secondary multimedia distribution.

## 1.4 Analysis and Debate

Although simply defined, cognitive radio concept features are much complex and hard to examine and implement. That is and as seen previously, few contributions attempt to consider both interactions between primary and secondary infrastructures and conflicts between the competing secondary users in addition to the shadowing and noise across the subchannels. Cognitive Radio networks are multi-hop, multichannel, multi-rate and multi-application. Thereby, characterizing reliability in CR networks is one of the major bottlenecks in the performance evaluation of multimedia traffic transmission in a secondary use scenario [23].

Several papers assume that all secondary users have perfect knowledge of their environment with no hidden nodes and when the primary user is off the air. They consider a perfect spectrum monitoring and ideal sensing operations. However, these are not realistic assumptions in a radio environment that may not be reciprocal (such as all FDD channels) and when the interferer cannot hear the interfered [24] [25]. Author should recognize the unsolved problem of "hidden nodes" which cannot be sensed and should consider the presence of a distributive mechanism for information exchange to facilitate the exploration of the surrounding environment or a centralized scheduler to collect the measurements and broadcast feedbacks.

## 1.5 Contribution

Currently, we are addressing the multimedia traffic transmission problem through Cognitive Radio networks (Fig .2). Here, we consider the primary user applications that have a traffic which can be modelled using the Markovian or the Poissonian processes (Fig. 3 and 4). Primary traffic parameters could be estimated based on a real observation window and using a maximum likelihood estimator for the Markov chain transition matrix $P^s$ and the initial distribution $\sigma^s$ in case of the Markovian traffics and the Poissonian arrival rate parameter $\lambda$ for the Poissonian primary arrivals. After sending the message, the used spectrum bands are sensed and the SUL will be restructured in case some packets got lost as a consequence of the PU appearance.

For ease of modelling, we adopt the TDMA method as a network sharing technique. The secondary communication operates in a slotted TDMA mode; the SUs transmit one after the other, each mostly using its specific time slot $i$ with some probability $q$. However, the same SU can transmit over the other slots if it has data to transmit out of its own slot, let $p$ be the probability that this SU transmits in the remaining slots $j \neq i$ (Fig. 2) [15].

The transmission performance on the proposed network model, as the realistic case, is mainly affected by two major factors: (1) interferences caused by the primary traffic arrival which leads to more corrupted SU packets, and (2) collisions caused by the fact that each SU could transmit opportunistically in other portions of time reserved for other SUs.

The source stream is progressively encoded and LT codes [16] are used to cope with packet losses caused by Primary User interference and other channel conditions.

For information exchange and distribution and similar to [18], we assume the existence of a trusted third party called centralized scheduler and we suggest to incorporate more self management capabilities and autonomous functions in Cognitive Radio devices. This creates a kind of a distributed cooperative system. The fundamental roles of the common scheduler are collecting different information communicated by secondary users like: sensing information, jammed subchannels, new acquired subchannels and subchannels characteristics whereas it provides feedback on that basis in terms of: hidden nodes and Secondary User Links assignment. These information flows up and down are exchanged using dedicated control channel, called Group Control Channel (GCC) [17]. Outlines of those protocols are given later.

Some graphs will be computed and plotted for some system parameters to describe the available trade-offs for a given multimedia transmission. We analyse the Spectral Efficiency in both primary traffic cases. We state that depending on the experienced primary traffic pattern parameters, the cognitive transmission behaves differently, so that we have to decide on the primary traffic pattern the most convenient for a given secondary transmission.

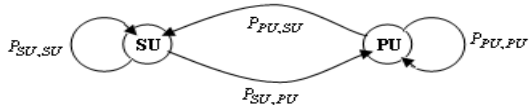

Fig. 3 Subchannel state model using Markov Chain

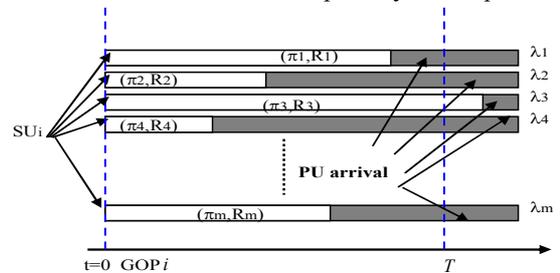

Fig. 4 Subchannel access model following a Poissonian process

This paper is organized as follows: Section 3 computes the analytic expression of the achieved Spectral Efficiency which considers the Primary traffic interruptions and the secondary collisions. In Section 4 we present the numerical results and we show the resulting gains in terms of system Spectral Efficiency, and finally Section 5 draws our conclusions.

## 2. Proposed Network Model

In this section, we will evaluate the multimedia traffic performance in a CR context. For this purpose, we give an analytic expression of the Spectral Efficiency metric which quantifies the optimal use of spectral resources.

### 2.1 General Analysis

We consider an infrastructure based CR network collocated with a licensed network. The primary traffic evolves following a given statistical law and Secondary Users $(SU_i)_{1 \leq i \leq M}$ share the temporarily unoccupied spectrum holes using the Opportunistic Spectrum Sharing principle. Throughout the paper, our main interest lies in introducing a model that customize the environmental parameters of the CR network and tries to encompass the chief factors shaping the performance of a secondary multimedia transmission.

Toward this end, we consider delay video transmission applications. We have opted for this choice regarding the fact that the video is very exigent in terms of required bandwidth. We assume that the video is source encoded using a layered compression scheme like SPIHT or MPEG. The video data consists of a group of pictures (GOP) each of $K$ packets. $L$ is the packet size.

For ease of exposition, the primary and secondary users are assumed to access the subchannels following a synchronous slot structure mode. In a TDMA based CR network, we investigate the problem of sharing idle time slots. The TDMA frame $T$ consists of $M$ slots each of the same time duration $T_{slot}$. At the start of every frame $T$ and during the sensing phase duration $T_{sens}$, all users conduct local spectrum sensing simultaneously to determine the unoccupied subchannels from which each $SU_i$ selects $S$ vacant subchannels to establish a secondary user link. Then, $SU_i$ starts transmitting its GOP packets over this link during the data duration $T_{data}$. So, we have $T = T_{sens} + T_{data}$.

Each cognitive device $SU_i$ always transmits in its assigned slot $i$ with probability $q$ and transmits with probability $p$ in the remaining time slots ($M-1$ slots) [15] [22].

We make use of LT codes to protect the multimedia traffic against PU interferences and also to mitigate the harmful collisions effects. Note $N$ as the number of LT encoded packets needed to recover the original $K$ transmitted packets with probability $1-DEP$. Based on simulations, an overhead of five percent can achieve low decoding error probability values [27] [29], so $N = 1.05 \times K$.

Although simply defined, cognitive radio concept features are much complex and hard to examine and implement. In practice, the transmission reliability over CR networks is influenced by many environmental parameters. In our scenario, we attempt to summarize those factors into two critical events. A secondary user succeeds its transmission if (1) for the secondary receiver, at least $N$ packets are received successfully from the set of selected subchannels $S$, and (2) the collisions with the surrounding neighbors remain at a relatively moderate or low level.

Let $u$ and $v$ be two active secondary users. The present paper aims at studying the Spectral Efficiency of the communication $u \rightarrow v$. This transmission is perturbed by the PU reclaims and the collisions between the concurrent transmissions.

Define $P_{success}$ as the transmission success probability of the communication $u \rightarrow v$ which takes into consideration only the primary interruptions and let $P_{collision}$ be the probability of collisions occurrence between the competing cognitive devices.

The successful transmission probability can be expressed as:

$$P_{u \rightarrow v} = P_{success} \times P_{no\ collision} \qquad (1)$$

We should notice that $P_{collision}$ does not depend on the primary traffic type since the concurrent access of secondary subscribers to the shared spectral resources is the same regardless of the primary traffic pattern.

### 2.2 An Analytical Expression for $P_{success}$

$P_{success}$ should be written as :

$$P_{success} = \Pr(N_T \geq N) \qquad (2)$$

Where $N_T$ is given by:

$$N_T = \sum_{s=1}^{S} N_s \tag{3}$$

$N_s$ denotes the number of packets transmitted over the subchannel $s$ with $s \in \{1, \ldots, S\}$.

Let each SC $s$ has a loss probability of fading and noise $\pi_s$ and channel capacity $R_s$. We suppose that the subchannel capacity is the same for all the subchannels, we note $R_0$ this capacity. Let $W$ be the subchannel bandwidth, the same for all subchannels.

The random variable $N_s$ is proportional to the available time on the $s$'th subchannel (Fig. 4).

Let $T_s^M$ be a random variable that denotes the available time on the subchannel $s$ under the Markovian primary traffic arrival and $T_s^P$ is the one corresponding to the Poissonian case.

## 2.3 Analytical Expression for $T_s^M$

Primary user traffic is modeled as a Markov process and the communication is assumed to be slotted then $T_s^M$ is given by the number of slots during which the subchannel $s$ remains free.

Similar to [33] and [34], we have:
$$T_s^M = mT_{slot} \tag{4}$$

Where the $m$'th slot is defined as the slot where the subchannel $s$ will change its state and becomes occupied by the corresponding PU, in other words the subchannel state remains vacant during the first $m$ slots and the PU captures its subchannel at the $(m+1)$'th slot.

Let $X_i^s$ denotes a random variable which represents the state of the subchannel $s$ at the $i$'th time slot where $i \in \{0, \ldots, T/T_{slot}\}$, $i = 0$ corresponds to the state of the subchannel $s$ just before the beginning of the transmission phase (subchannel sensing state).

The distribution $X_i^s$ is considered as a continuous time homogeneous Markov chain process determined by the initial distribution $\sigma^s$ and the transition probabilities matrix $P^s$.

The finite set of states of $X_i^s$ is defined by:
$$G = \{SU, PU\} \tag{5}$$

Where $SU$ refers to the state of the subchannel when it is not in use by the primary traffic i.e. ready for secondary use, and $PU$ represents the subchannel state when it is busy.

Using Eq. (5), we get:
$$\sigma^s = \begin{pmatrix} \Pr(X_0^s = SU) \\ \Pr(X_0^s = PU) \end{pmatrix} \tag{6}$$

And:
$$P^s = \begin{pmatrix} P_{SU,SU}^s & P_{SU,PU}^s \\ P_{PU,SU}^s & P_{PU,PU}^s \end{pmatrix} \tag{7}$$

We may notice that both events $\{T_s^M = mT_{slot}\}$ and $\{X_{m+1}^s = PU \cap X_m^s = SU \cap \cdots \cap X_0^s = SU\}$ are equivalent.

Therefore, we could proceed in the following manner:
For $m = 0$:
$$\Pr(T_s^M = 0) = \Pr(\{X_1^s = PU \cap X_0^s = SU\} \cup \{X_0^s = PU\}) \tag{8}$$

And for $m \in \{1, \ldots, T/T_{slot} - 1\}$:
$$\begin{aligned} &\Pr(T_s^M = mT_{slot}) = \\ &\Pr(X_{m+1}^s = PU \cap X_m^s = SU \cap \cdots \cap X_0^s = SU) \end{aligned} \tag{9}$$

Finally for $m = T/T_{slot}$ we have:
$$\begin{aligned} &\Pr(T_s^M = T) = \\ &\Pr(X_{T/T_{slot}}^s = SU \cap X_{T/T_{slot}-1}^s = SU \cap \cdots \cap X_0^s = SU) \end{aligned} \tag{10}$$

Using the Markov chain properties, we develop the equation Eq. (9) as:
$$\begin{aligned} &\Pr(X_{m+1}^s = PU \cap X_m^s = SU \cap \cdots \cap X_0^s = SU) = \\ &\sigma^s(1) \times \underbrace{P_{SU,SU}^s \times \cdots \times P_{SU,SU}^s}_{m \text{ times}} \times P_{SU,PU}^s \end{aligned} \tag{11}$$

And then,
$$\begin{aligned} &\Pr(X_{m+1}^s = PU \cap X_m^s = SU \cap \cdots \cap X_0^s = SU) = \\ &\sigma^s(1) \times \left(P_{SU,SU}^s\right)^m \times P_{SU,PU}^s \end{aligned} \tag{12}$$

Equations Eq. (8) and Eq. (10) should be expressed in the same way.

Let $\gamma_s$ be the threshold level that minimizes the spectrum sensing error probability on the subchannel $s$ [18], it is essentially important for the SU to design the optimal sensing threshold [19]. Regarding the fact that during the sensing phase the subchannels selected to set up a SUL should be judged as spectrum holes to a meaningful degree, $\sigma^s$ would be given by:
$$\sigma^s = \begin{pmatrix} \gamma_s \\ 1 - \gamma_s \end{pmatrix} \tag{13}$$

Hence and from Eq. (8), Eq. (10), Eq. (12) and Eq. (13), the probability density function (PDF) of $T_s^M$ can be evaluated by:

$$PDF(T_s^M) = \begin{cases} \gamma_s \times P_{SU,PU}^s + 1 - \gamma_s, & \text{if } T_s^M = 0 \\ \gamma_s \times \left(P_{SU,SU}^s\right)^m \times P_{SU,PU}^s, & \text{if } T_s^M = mT_{slot} \\ & 1 \leq m \leq T/T_{slot} - 1 \\ \gamma_s \times \left(P_{SU,SU}^s\right)^{T/T_{slot}}, & \text{otherwise} \end{cases} \quad (14)$$

## 2.4 Analytical Expression for $T_s^P$

Primary user traffic is modeled as a Poisson process then $T_s^P$ is given by:

$$T_s^P = \begin{cases} \tau_s, & \text{if } \tau_s \leq T \\ T, & \text{if } \tau_s \succ T \end{cases} \quad (15)$$

Where $\tau_s \sim \exp(\lambda_s)$

## 2.5 Analytical Expression for $N_s$

Consequently, $N_s$ could be represented as:
(1) Markovian case,

$$N_s = \frac{(1-\pi_s) \times R_0 \times T_s^M}{T} \quad (16)$$

(2) Poissonian case,

$$N_s = \frac{(1-\pi_s) \times R_0 \times T_s^P}{T} \quad (17)$$

With $T_s^M$ and $T_s^P$ defined by the distribution given, respectively, in Eq. (14) and Eq. (15).

In fact, we have entirely defined the random variables $(N_s)_{s \in \{1,...,S\}}$. Therefore, using the equations Eq. (2), Eq. (3), Eq. (16) and Eq. (17) and the property that the PDF of a sum of Random Variables is computed as the convolution of the PDFs of those variables i.e. formula Eq. (18), we can compute $P_{success}$.

$$PDF(N_T) = \otimes_{s \in \{1,...,S\}} PDF(N_s) \quad (18)$$

## 2.6 An Analytical Expression for $P_{collision}$

Let $Deg_v$ be the number of neighbors of the active Cognitive user $v$. Remember that $q$ is the probability that $u$ transmits in its assigned time slot $i$ and $p$ the probability that it transmits in the remaining time slots $j \neq i$ [15].

If we consider several structured SULs which are pairwise disjoin as defined in [15], the total average probability of collisions over the available Secondary User Links $N_{sul}$ is:

$$P_{collision} = \left(1 - \frac{q(1-p) + (M-1)p(2-p-q)}{M}(1-p)^{Deg_v-1}\right)^{N_{sul}} \quad (19)$$

## 2.7 An Analytical Expression for Spectral Efficiency

As in [12], the total Spectral Efficiency can be computed as:

$$SE = \frac{(1-DEP) \times (1-P_{collision}) \times P_{success} \times K \times L}{S \times W \times T} \quad (20)$$

## 2.8 Cross layer functionalities in CR networks

The number of slots $M$ is fixed by the centralized scheduler according to the number of surrounding cognitive peers and each cognitive device $v$ broadcasts the number of its neighbors $Deg_v$. The sensing stage duration $T_{sens}$ should be stated in such a manner as to ensure a good compromise between sensing performance, throughput [30] [31] and data transmission [32]. During the sensing phase, the CR system determines the vacant subchannels ready for secondary use using a cooperative distributed system constituted by the cognitive devices in addition to the centralized peer. Such distributed infrastructures are very suitable for real time systems like CR networks and enable reducing the complexity of the numerical operations and data processing needed for ensuring CR features. The flow of information comprises all the vital information like the jammed subchannels, subchannels characteristics like shadowing and noise, traffic observations necessary for traffic patterns estimation and false alarm ad miss detection probabilities estimation.

Once the necessary information is collected with respect to the available time for the sensing stage, the centralized scheduler proceeds to analyzing the collected information and measurements in order to infer the most perfect picture of the surrounding environment and the closer to the reality. Afterwards, the centralized scheduler should run some algorithms like the one introduced in [15] serving as a mean of creation of many SULs ( $N_{sul}$ ) having high Spectral Efficiency. Accordingly, each SU is assigned a specific secondary user link.

Using the collected primary traffic observations, the centralized scheduler should make a decision about the primary traffic patterns via the maximum likelihood estimator and then it can chooses the primary traffic that suits well the given transmission media data rate in case of the existence of many primary patterns.

The use of a cooperative sensing system should alleviate the occurrence of many recurrent problems such as the hidden node issue. For clarity, the hidden node problem is where a transmitter/receiver pair of secondary users is in the shadow of a primary user, perhaps behind a building or hill or truck, and so their sensing does not detect the primary user. Since they don't detect any Primary transmissions in the band and in those locations, they commence to transmit, even though the primary user may have a receiver (say a TV set or pocket pager) that is outside the shadow of the building or hill or truck, but still in range of the secondary cognitive radios. So it is interfered with, perhaps immensely so.

The average traffics $p$ and $q$ could be analytically estimated by addressing the Spectral Efficiency balance described in [15] and then those optimal parameters should broadcasted to all secondary subscribers.

At the end of the frame, the centralized scheduler should capitalize on the current transmission experiences to take advantages of the factors that promote and/or penalize the transmission reliability. Previous knowledge and experiences could be reused to avoid repetitive operations and calculations. In the last case, the centralized scheduler must store the static parameters or the parameters that are not varying fast like subchannels shadowing coefficients in a specific internal database. Details of storage issue are out of scope of the current work.

Secondary users have a huge potential to be a valuable part of the value chain. The centralized scheduler could delegate some tasks to SUs in case they are inactive and dormant.

## 3. Numerical Results

For numerical simulations, consider a cognitive radio network where a cognitive peer $u$ is conveying a GOP of $K = 3000$ packets to a cognitive destination $v$ over some vacant spectrum rooms. Unless otherwise stated, the slots number is $M = 10$, the cognitive user $v$ has three neighbors $Deg_v = 3$ and we assume that the execution of the duplication-based algorithm defined in [15] results in five available SULs $N_{sul} = 5$. The packet size is about $L = 1000\,bits$.

The available pool of subchannels has a size of 9, a common subchannel capacity of $R_0 = 10\,Mbps$ and a subchannel bandwidth of $W = 100\,kHz$.

Markovian primary users evolve according to the following set of parameters:
$P_{SU,SU} = [0.9\ 0.7\ 0.76\ 0.8\ 0.68\ 0.82\ 0.77\ 0.65\ 0.45]$

Let's take the following noise and fading losses distribution:
$\pi = [0.03\ 0.04\ 0.01\ 0.02\ 0.05\ 0.025\ 0.06\ 0.01\ 0.03]$
The time frame duration is about $T = 1\,s$ and the sensing duration is set to $T_{sens} = 5\,ms$.

For LT coding and decoding processes, the Robust Soliton distribution has as parameters $c = 0.1$ and $\delta = 0.5$, we consider a low *DEP* value of 1%.

The estimated traffic average on the assigned slot is $q = 90\%$.

For fair comparison, $\gamma_s$ is assumed to be the same for all the subchannels pool: $\gamma_s = 1$.

In Fig. 5, the computed $P_{success}$ is plotted against the subchannels number $S$. The Poissonian primary traffic accesses the subchannels following this set of parameters: $\lambda = [3\ 2\ 1\ 2.5\ 3.6\ 4\ 6\ 2.4\ 3.2]$. It can be observed that the Poissonian primary traffic performs better than the Markovian pattern. In fact, in the Poissonian case, if a certain type of event occurs on average of $J$ times per period $T$, to analyze the number of events occurring in this period we choose as model a Poisson distribution with parameter $\lambda = J \times T$. Thus, for low $\lambda$ values, the Poissonian primary users utilize their licensed bands intermittently and the Poissonian primary traffic arrival happening $J$ becomes more infrequent. So that, more chance that the subchannel remains idle during $T$.

Fig. 6 shows the achieved probability $P_{success}$ in terms of subchannels number $S$ for high $\lambda$ values: $\lambda = [30\ 20\ 10\ 25\ 36\ 40\ 60\ 24\ 32]$. It is obvious that the Markovian primary traffic surpasses the Poissonian one for this range of $\lambda$ values. Using the previously stated mathematical relationship $\lambda = J \times T$, the Poissonian primary reclaims get increased when increasing the Poissonian coefficient $\lambda$ since it becomes very likely that the primary user captures its corresponding subchannel.

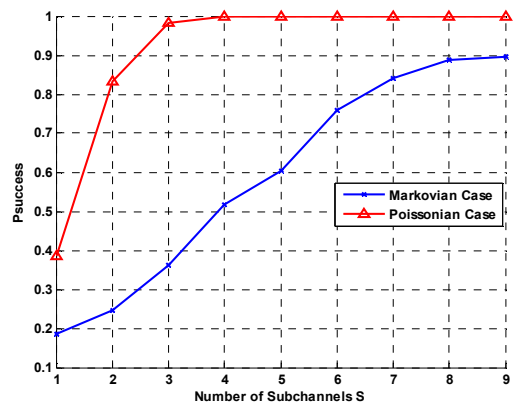

Fig. 5 Probability $P_{success}$ versus number of subchannels $S$

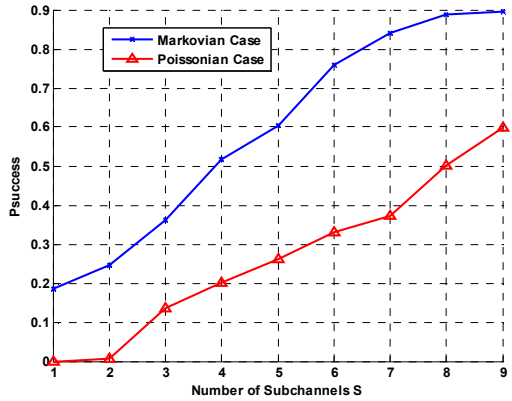

Fig. 6 Probability of successfully receiving more than $N$ packets versus number of subchannels $S$

Fig. 7 depicts the impact of moderate values of $\lambda$ on the total achieved success probability $P_{success}$ plotted against the number of free subchannels $S$. The parameter $\lambda$ has been fixed to $[18\ 12\ 6\ 15\ 21.6\ 24\ 36\ 14.4\ 19.2]$. It is clearly seen that the Markovian case outperforms the Poissonian one in both ranges $[1, 3]$ and $[6, 8]$. Nevertheless, it is the other way round outside the given ranges. The secondary user link size $S$ becomes a decisive factor to choose the primary traffic pattern we deem best for the given media data rate.

In Fig. 8, we examine the impact of the parameter $\lambda$ on the Spectral Efficiency $SE$ computed for several $p$ values.

For $\lambda = [3\ 2\ 1\ 2.5\ 3.6\ 4\ 6\ 2.4\ 3.2]$, the Poissonian primary traffic performs better efficacy compared to the Markovian arrival process. This is due to the fact that the less the Poissonian coefficient $\lambda$, the more the probability $P_{success}$, and then the more the Spectral Efficiency $SE$. It is also interesting to note that, there is some $p$ value that maximizes the achieved Spectral Efficiency ($p \approx 0.2$).

Finally, Fig. 9 illustrates the achieved $SE$ metric in terms of average traffic $p$ for high $\lambda$ values. Simulations were run for $\lambda = [30\ 20\ 10\ 25\ 36\ 40\ 60\ 24\ 32]$. The Markovian primary traffic yields better results than the Poissonian distribution. Obviously, high $\lambda$ values means that more subchannels can be subject to eventual primary interruptions and those prematurely captured subchannels degrade considerably the effectiveness of the utilization of the spectrum. On the other hand, high traffic performance is attained where approaching the optimal value of $p \approx 0.2$ and the Spectral Efficiency approaches its maximum value $SE = 1.65$ for the Markovian case and $SE = 1.1$ for the Poissonian one.

## 4. Conclusions

The present paper tackles the issue of multimedia traffic distribution in TDMA-based Cognitive Radio networks with the assumption that the primary system evolves following either a Markovian process or a Poissonian distribution. The impact of the primary traffic interruptions on the secondary traffic has been examined and we used a general model for collisions to modelize the opportunistic access of secondary users to shared rooms. Numerical results have conduced a comparative analyze between the two traffic types depending upon the network state. The obtained findings emphasize the complexity of the dynamics between different peers in CR networks and outline the necessity of describing the available trade-offs to be able to choose the optimal balance of the system.

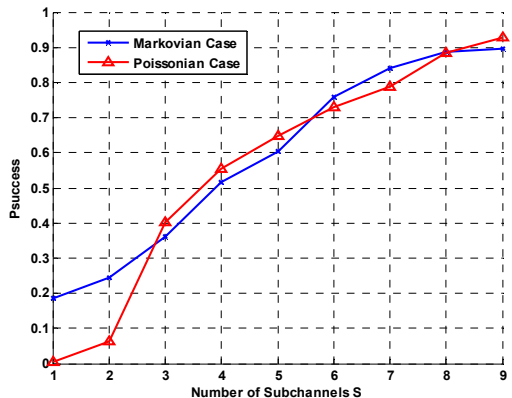

Fig. 7 Probability $P_{success}$ versus number of subchannels $S$

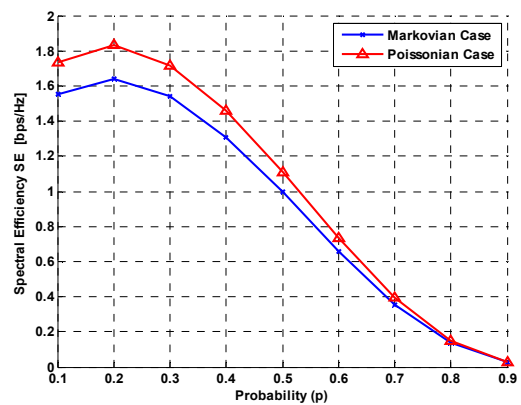

Fig. 8 Spectral Efficiency comparison for different values of $p$

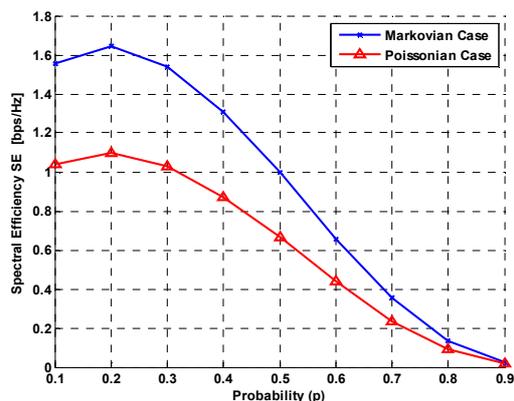

Fig. 9 Spectral Efficiency versus traffic average $p$ for both primary traffic patterns

## References

[1] http://www.fcc.gov/oet/info/database/spectrum/
[2] Shared Spectrum Compagny. Spectrum occupancy measurement. site internet, http://www.sharedspectrum.com/measurements/.
[3] NTIA, "U.S. frequency allocations." [Online]. Available: http://www.ntia.doc.gov/osmhome/allochrt.pdf
[4] J. Mitola III, "Cognitive Radio: An Integrated Agent Architecture for Software Defined Radio," Ph.D Thesis, KTH Royal Institute of Technology, 2000.
[5] FCC, ET Docket no. 03-322. Notice of Proposed Rule Making and Order, December 2003.
[6] D. Cabric, S. M. Mishra, D. Willkomm, R. W. Broderson, and A. Wolisz, "A cognitive radio approach for usage of virtual unlicensed spectrum," in 14th IST Mobile Wireless Communications Summit 2005, Dresden, Germany, June 2005.
[7] A. S. Kamil and I. Khider, "Open Research issues in Cognitive Radio," Proc. 16th Telecommunications forum TELFOR, pp. 250–253, Nov. 25-27, 2008.
[8] Ben Letaief, K.; Wei Zhang; , "Cooperative Communications for Cognitive Radio Networks," Proceedings of the IEEE , vol.97, no.5, pp.878-893, May 2009.
[9] T. Weiss and F. Jondral, "Spectrum Pooling: An Innovative Strategy for the Enhancement of Spectrum Efficiency", IEEE Communications Magazine, Vol. 42, no. 3, March 2004, pp. 8-14.
[10] H. Kushwaha, Y. Xing, R. Chandramouli, and K. P. Subbalakshmi, "Erasure Tolerant Coding for Cognitive Radios", Book Chapter in Cognitive Networks: Towards Self-Aware Networks, Ed. Qusay H. Mahmoud, Wiley, Expected 2007.
[11] A. Chaoub, E. Ibn Elhaj, and J. El Abbadi, "Multimedia traffic transmission over Cognitive Radio networks using Multiple Description Coding," ACC 2011, Part I, CCIS 190, pp. 529–543, 2011, Springer-Verlag Berlin Heidelberg 2011.
[12] H. Kushwaha, Y. Xing, R. Chandramouli, and H. Heffes, "Reliable multimedia transmission over cognitive radio networks using fountain codes," Proc. IEEE, vol. 96, no. 1, pp. 155–165, Jan. 2008.
[13] Xin-Lin Huang; Gang Wang; Fei Hu; Kumar, S.; , "The Impact of Spectrum Sensing Frequency and Packet-Loading Scheme on Multimedia Transmission Over Cognitive Radio Networks," Multimedia, IEEE Transactions on , vol.13, no.4, pp.748-761, Aug. 2011.
[14] I. Demirkol, C. Ersoy, F. Alagoz, and H. Delic, "The impact of a realistic packet traffic model on the performance of surveillance wireless sensor networks," Computer Networks, vol. 53, no. 3, pp. 382-399, 2009.
[15] A. Chaoub, E. Ibn Elhaj, and J. El Abbadi, "Multimedia traffic transmission over TDMA shared cognitive radio networks with poissonian primary traffic," Multimedia Computing and Systems, 2011. ICMCS '11. International Conference on , vol., no., pp.378-383, 7-9 April 2011.
[16] M. Luby, "LT codes", Proc. 43rd Ann. IEEE Symp. on Foundations of Computer Science, 2002, pp. 271–282.
[17] D. Willkomm, J. Gross, and A. Wolisz, "Reliable link maintenance in cognitive radio systems," in Proc. IEEE Symp. New Frontiers Dyn. Spectrum Access Netw. (DySPAN 2005), Baltimore, MD, Nov. 2005.
[18] A. Chaoub, E. Ibn Elhaj, and J. El Abbadi, "Video transmission over cognitive radio TDMA networks under collision errors," International Journal of Advanced Computer Science and Application (IJACSA), (2011), Special Issue on Wireless and Mobile Networks, August 2011, pp.5-13.
[19] A. Chaoub, E. Ibn Elhaj, and J. El Abbadi, "Binomial and poissonian primary traffics in cognitive radio networks," 2nd Int. Workshop on Codes, Cryptography and Communication Systems "WCCCS'11", Rabat, June 16-17, 2011, pp 238-243.
[20] Q. Zhao, L. Tong, and A. Swami, "Decentralized cognitive MAC for dynamic spectrum access," in Proc. IEEE DySPAN 2005, Baltimore, MD, 2005.
[21] R. Urgaonkar and M.J. Neely. "Opportunistic scheduling with reliability guarantees in cognitive radio networks," In INFOCOM 2008. The 27th Conference on Computer Communications. IEEE, pages 1301-1309, April 2008.
[22] L. Cuiran and L. Chengshu , "Opportunistic spectrum access in cognitive radio networks," Neural Networks, 2008. IJCNN 2008. (IEEE World Congress on Computational Intelligence). IEEE International Joint Conference on , vol., no., pp.3412-3415, 1-8 June 2008.
[23] Pal, R.; Idris, D.; Pasari, K.; Prasad, N.; , "Characterizing reliability in cognitive radio networks," Applied Sciences on Biomedical and Communication Technologies, 2008. ISABEL '08. First International Symposium on , vol., no., pp.1-6, 25-28 Oct. 2008.
[24] I. F. Akyildiz, W.-Y. Lee, M. C. Vuran, and S. Mohanty, "Next generation/dynamic spectrum access/cognitive radio wireless networks: A survey," Comput. Netw., vol. 50, pp. 2127–2159, May 2006.
[25] D. Cabric, S. M. Mishra, and R. W. Brodersen, "Implementation issues in spectrum sensing for cognitive radios," in Proc. Asilomar Conf. Signals, Syst., Comput., Nov. 7–10, 2004, vol. 1, pp. 772–776.

[26] R. W. Broderson, A. Wolisz, D. Cabric, S. M. Mishra, and D. Willkomm, "Corvus: A cognitive radio approach for usage of virtual unlicensed spectrum," White Paper, Univ. California Berkeley, Tech. Rep., Jul. 2004.

[27] D.J.C. MacKay, "Fountain codes", IEE Proc.-Commun., vol. 152(6), 2005, pp.1062-1068.

[28] I.S. Reed and G. Solomon, "Polynomial codes over certain finite fields," J. Soc. Ind. Appl. Math. vol. 6, June 1960, pp. 16-21.

[29] C. R. Berger, S. Zhou, Y. Wen, K. Pattipati, and P. Willett, "Optimizing joint erasure- and error-correction coding for wireless packet transmissions," IEEE Trans. Wireless Commun., vol. 7, no. 11, pp. 45864595, Nov. 2008.

[30] Ying-Chang Liang; Yonghong Zeng; Peh, E.C.Y.; Anh Tuan Hoang; , "Sensing-Throughput Tradeoff for Cognitive Radio Networks," Wireless Communications, IEEE Transactions on , vol.7, no.4, pp.1326-1337, April 2008.

[31] Penna, F.; Khaleel, H.; Pastrone, C.; Tomasi, R.; Spirito, M.; , "On spectrum sensing duration in cognitive wireless sensor networks," Applied Sciences in Biomedical and Communication Technologies (ISABEL), 2010 3rd International Symposium on , vol., no., pp.1-5, 7-10 Nov. 2010.

[32] Yulong Zou; Yu-Dong Yao; Baoyu Zheng; , "Spectrum sensing and data transmission tradeoff in cognitive radio networks," Wireless and Optical Communications Conference (WOCC), 2010 19th Annual , vol., no., pp.1-5, 14-15 May 2010.

[33] A. Chaoub, E. Ibn-Elhaj, "Markovian primary traffics in Cognitive Radio networks," Electrical and Control Engineering (ICECE), 2011 International Conference on , vol., no., pp.5987-5991, 16-18 Sept. 2011.

[34] A. Chaoub, E. Ibn Elhaj, "Modelling Multimedia Traffic over Cognitive Radio Networks using Markov Chain under Collision Errors," International Journal of Wisdom Based Computing (IJWBC), Vol. 1, 2011, pp.139-145.